\documentclass[useAMS,usenatbib]{mn2e}
\usepackage{times}
\usepackage{graphicx}
\usepackage{subfigure}
\usepackage{psfrag}
\usepackage{amsmath}
\usepackage{amssymb}

\def\reff@jnl#1{{\rm#1\/}}
\def\aj{\reff@jnl{AJ}}                 
\def\araa{\reff@jnl{ARA\&A}}           
\def\apj{\reff@jnl{ApJ}}               
\def\apjl{\reff@jnl{ApJ}}              
\def\apjs{\reff@jnl{ApJS}}             
\def\ao{\reff@jnl{Appl.Optics}}        
\def\apss{\reff@jnl{Ap\&SS}}           
\def\aap{\reff@jnl{A\&A}}              
\def\aapr{\reff@jnl{A\&A~Rev.}}        
\def\aaps{\reff@jnl{A\&AS}}            
\def\azh{\reff@jnl{AZh}}               
\def\baas{\reff@jnl{BAAS}}             
\def\jrasc{\reff@jnl{JRASC}}           
\def\memras{\reff@jnl{MmRAS}}          
\def\mnras{\reff@jnl{MNRAS}}           
\def\pra{\reff@jnl{Phys.Rev.A}}        
\def\prb{\reff@jnl{Phys.Rev.B}}        
\def\prc{\reff@jnl{Phys.Rev.C}}        
\def\prd{\reff@jnl{Phys.Rev.D}}        
\def\prl{\reff@jnl{Phys.Rev.Lett}}     
\def\pasp{\reff@jnl{PASP}}             
\def\pasj{\reff@jnl{PASJ}}             
\def\qjras{\reff@jnl{QJRAS}}           
\def\skytel{\reff@jnl{S\&T}}           
\def\solphys{\reff@jnl{Solar~Phys.}}   
\def\sovast{\reff@jnl{Soviet~Ast.}}    
\def\ssr{\reff@jnl{Space~Sci.Rev.}}    
\def\zap{\reff@jnl{ZAp}}               
\def\nat{\reff@jnl{Nature}}            

\title[Weak lensing by triaxial galaxy clusters]{Weak lensing by triaxial galaxy clusters} 
\author[F. Feroz and M.P. Hobson] 
{F.~Feroz\thanks{E-mail: f.feroz@mrao.cam.ac.uk} and M.~P.~Hobson\\ 
Astrophysics Group, Cavendish Laboratory, JJ Thomson Avenue, Cambridge CB3 0HE, UK\\}

\date{Accepted ---. Received ---; in original form 11 August 2011}
\pagerange{\pageref{firstpage}--\pageref{lastpage}}
\pubyear{2010}

\begin{document}
\label{firstpage}
\maketitle

\begin{abstract}
Weak gravitational lensing studies of galaxy clusters often assume a
spherical cluster model to simplify the analysis, but some recent
studies have suggested this simplifying assumption may result in large
biases in estimated cluster masses and concentration values, since
clusters are expected to exhibit triaxiality. Several such analyses
have, however, quoted expressions for the spatial derivatives of the
lensing potential in triaxial models, which are open to
misinterpretation. In this paper, we give a clear description of weak
lensing by triaxial NFW galaxy clusters and also present an efficient
and robust method to model these clusters and obtain parameter
estimates. By considering four highly triaxial NFW galaxy clusters, we
re-examine the impact of the simplifying assumption of sphericity and
find that while the concentration estimates are largely unbiased,
except in one of our triaxial NFW simulated clusters, the masses are
significantly biased, by up to $40\%$, for all the clusters we
analysed. Moreover, we find that erroneously assuming spherical
symmetry can lead to the mistaken conclusion that some substructure is
present in the galaxy clusters or, even worse, that multiple galaxy
clusters are present in the field. Our cluster fitting method also
allows one to answer the question of whether a given cluster exhibits
triaxiality or a simple spherical model is good enough.
\end{abstract}


\section{Introduction}\label{sec:intro}

Clusters of galaxies are the most massive gravitationally bound
objects in the universe and, as such, are critical tracers of the
formation of large-scale structure. The number counts of clusters as a
function of their mass and redshift have been predicted both
analytically (see e.g. \citealt{press74,sheth01}) and from large scale
numerical simulations (see e.g. \citealt{jenkins01,evrard02}), and are
particularly sensitive to the cosmological parameters $\sigma_8$ and
$\Omega_{\rm m}$ (see e.g. \citealt{battye03}). The size and formation
history of massive clusters is such that the ratio of gas mass to
total mass is expected to be representative of the universal ratio
$\Omega_{\rm b}/\Omega_{\rm m}$, once the relatively small amount of
baryonic matter in the cluster galaxies is taken into account (see
e.g. \citealt{white93}).

The study of cosmic shear has progressed rapidly in recent years, with
the availability of high quality wide-field lensing data. Large
dedicated surveys with ground-based telescopes have been employed to
reconstruct the mass distribution of the universe and constrain
cosmological parameters (see e.g. \citealt{massey07, massey05,
  hoekstra06}). Weak lensing also allows one to detect galaxy clusters
without making any assumptions about the baryon fraction, richness,
morphology or dynamical state of the cluster, and so weak lensing
cluster modelling would allow one to test these assumptions by
observing the cluster with optical, X-ray or Sunyaev--Zel'dovich (SZ)
methods. Nonetheless, weak lensing studies are very challenging, not
only because the data are very noisy, but also because of the inherent
under-determined nature of the problem, namely constraining the 3-D
structure of the galaxy cluster using lensing observations that are
sensitive only to the 2-D projected mass distribution, at least in the
thin lens approximation.

A model cluster density profile can be determined from numerical
$N$-body simulations of large-scale structure formation in a
$\Lambda$CDM universe. In particular, assuming spherical symmetry, the
NFW profile (\citealt{navarro97}) provides a good fit to the
simulations. The NFW profile is parameterized by the virial mass
$M_{\rm 200}$ and the concentration parameter $C$, which are
highly-correlated. The concentration parameter is directly related to
the central density of the cluster, which depends on cluster's
formation history. Since galaxy clusters are the most recently bound
objects in the universe, the cluster concentrations serve as probes of
the mean density of the universe at lower redshifts. $N$-body
simulations predict $C \sim 4$ for a cluster of mass $M_{\rm 200} =
10^{15} h^{-1} M_{\sun}$. Several studies, assuming a spherical NFW
model, have reported weak lensing results with very high concentration
values of $C \sim 8$ (see e.g. \citealt{2007ApJ...668..643L,
  2003ApJ...598..804K, 2003A&A...403...11G}), casting doubt on the
$\Lambda$CDM model. In order to shed further light on this
discrepancy, it is important to understand to what extent can these
results be explained by deviations from spherical symmetry. Another
reason for studying the triaxiality of galaxy clusters is that several
studies on structure formation in $\Lambda$CDM suggest clusters often
exhibit significant triaxiality (see e.g. \citet{bett07, shaw06}).

Several attempts have already been made towards understanding the
impact of triaxiality in explaining the deviations from $\Lambda$CDM
of the observed gravitational lensing
results. \citet{2002ApJ...574..538J} first presented a full
parameterization for a triaxial NFW halo. \citet{2005ApJ...632..841O}
and \citet{2005A&A...443..793G} showed that a triaxial NFW fit to
gravitational lensing observations of clusters Abell 1689 and
MS2137-23 yields parameter estimates consistent with $N$-body
simulation. \citet{corless07} and \citet{corless08} presented a
general Markov Chain Monte Carlo (MCMC) method for fitting a triaxial
NFW model to weak lensing observations. From the analysis of simulated
lensing observations of triaxial clusters, they concluded that
significant elongation of the cluster along the line of sight can
cause its mass to be overestimated by up to $50$ per cent and its
concentration by as much as a factor of $2$, if one simply assumes a
spherical NFW model when analysing its lensing signal.  This claim is
important since it would significantly ease the tension between the
$\Lambda$CDM model and the very high values of the concentration
parameter reported in a number of clusters, derived from lensing
observations assuming a spherical NFW model.

The triaxial lensing equations quoted in \citet{corless07} and
\citet{corless08} adopted the triaxial NFW model equations from
\citet{2001astro.ph..2341K}, which can easily be misinterpreted, and
also had some additional errors, although the correct form of
equations were used in their analysis code (King, private
communication). In any case, our purpose in this paper is to present a
clear description of the weak lensing equations for the triaxial NFW
model and simultaneously introduce an efficient, robust and accurate
method, based on nested sampling, for fitting triaxial NFW model to
weak lensing observations. In particular, we study the discrepancies
that may arise in fitting a spherical NFW model to a galaxy cluster
exhibiting significant triaxiality.

The outline of this letter is as follows. In Sec. \ref{sec:bayesian}
we introduce Bayesian inference, which we use for cluster fitting. In
Sec. \ref{sec:lensing} we describe the triaxial NFW model and derive
the relevant equations. We apply our method to a set of simulated weak
lensing simulations in Sec. \ref{sec:application}.  Finally we present
our conclusions in Sec. \ref{sec:conclusions}.

\section{Bayesian inference}\label{sec:bayesian}

Our cluster fitting methodology is built upon the principles of Bayesian inference, which provides a consistent approach to the estimation of a set of
parameters $\mathbf{\Theta}$ in a model (or hypothesis) $H$ for the data $\mathbf{D}$. Bayes' theorem states that
\begin{equation} 
\Pr(\mathbf{\Theta}|\mathbf{D}, H) = \frac{\Pr(\mathbf{D}|\,\mathbf{\Theta},H)\Pr(\mathbf{\Theta}|H)}{\Pr(\mathbf{D}|H)},
\end{equation}
where $\Pr(\mathbf{\Theta}|\mathbf{D}, H) \equiv P(\mathbf{\Theta})$ is the posterior probability distribution of the parameters,
$\Pr(\mathbf{D}|\mathbf{\Theta}, H) \equiv \mathcal{L}(\mathbf{\Theta})$ is the likelihood, $\Pr(\mathbf{\Theta}|H) \equiv \pi(\mathbf{\Theta})$ is the
prior, and $\Pr(\mathbf{D}|H) \equiv \mathcal{Z}$ is the Bayesian evidence.

In parameter estimation, the normalising evidence factor is usually ignored, since it is independent of the parameters $\mathbf{\Theta}$, and inferences are
obtained by taking samples from the (unnormalised) posterior using standard MCMC sampling methods, where at equilibrium the chain contains a set of samples
from the parameter space distributed according to the posterior. This posterior constitutes the complete Bayesian inference of the parameter values, and can
be marginalised over each parameter to obtain individual parameter constraints.

In contrast to parameter estimation problems, for model selection the evidence takes the central role and is simply the factor required to normalize the
posterior over $\mathbf{\Theta}$:
\begin{equation}
\mathcal{Z} = \int{\mathcal{L}(\mathbf{\Theta})\pi(\mathbf{\Theta})}\mathrm{d}^D\mathbf{\Theta},
\label{eq:3}
\end{equation} 
where $D$ is the dimensionality of the parameter space. As the average of the likelihood over the prior, the evidence is larger for a model if more of its
parameter space is likely and smaller for a model with large areas in its parameter space having low likelihood values, even if the likelihood function is
very highly peaked. Thus, the evidence automatically implements Occam's razor. The question of model selection between two models $H_{0}$ and $H_{1}$ can
then be decided by comparing their respective posterior probabilities given the observed data set $\mathbf{D}$, as follows
\begin{equation}
R = \frac{\Pr(H_{1}|\mathbf{D})}{\Pr(H_{0}|\mathbf{D})}
  = \frac{\Pr(\mathbf{D}|H_{1})\Pr(H_{1})}{\Pr(\mathbf{D}| H_{0})\Pr(H_{0})}
  = \frac{\mathcal{Z}_1}{\mathcal{Z}_0} \frac{\Pr(H_{1})}{\Pr(H_{0})},
\label{eq:3.1}
\end{equation}
where $\Pr(H_{1})/\Pr(H_{0})$ is the a priori probability ratio for the two models, which can often be set to unity but occasionally requires further
consideration.

Evaluation of the multidimensional integral in (\ref{eq:3}) is a challenging numerical task. Standard techniques like thermodynamic integration are extremely
computationally intensive which makes evidence evaluation at least an order of magnitude more costly than parameter estimation. Some fast approximate methods
have been used for evidence evaluation, such as treating the posterior as a multivariate Gaussian centred at its peak (see e.g. \citealt{hobson03}), but this
approximation is clearly a poor one for multimodal posteriors (except perhaps if one performs a separate Gaussian approximation at each mode). The
Savage-Dickey density ratio has also been proposed (see e.g. \citealt{trotta05}) as an exact, and potentially faster, means of evaluating evidences, but is
restricted to the special case of nested hypotheses and a separable prior on the model parameters. Various alternative information criteria for astrophysical
model selection are discussed by \citet{liddle07}, but the evidence remains the preferred method.

The nested sampling approach, introduced by \citet{skilling04}, is a Monte Carlo method targeted at the efficient calculation of the evidence, but also
produces posterior inferences as a by-product. \citet{feroz08} and \citet{multinest} built on this nested sampling framework and have introduced the
{\sc MultiNest} algorithm which is very efficient in sampling from posteriors that may contain multiple modes and/or large (curving) degeneracies and also
calculates the evidence. This technique has greatly reduced the computational cost of Bayesian parameter estimation and model selection, and is employed in
this paper.

\section{Weak lensing by Triaxial Galaxy Clusters}\label{sec:lensing}

\subsection{Weak lensing background}\label{sec:lensing:background}

A cluster mass distribution is investigated using weak gravitational lensing through the relationship (see e.g.  \citealt{schramm95})
$\langle\epsilon(\bmath{x})\rangle = g(\bmath{x})$, that is, at any point $\bmath{x}$ on the sky, the local average of the complex ellipticities $\epsilon =
\epsilon_1 + i \epsilon_2$ of a collection of background galaxy images is an unbiased estimator of the local complex reduced shear field, $g = g_1 + i g_2$,
due to the cluster. Adopting the thin-lens approximation, for a projected mass distribution $\Sigma(\bmath{x})$ in the lens, the reduced shear $g(\bmath{x})$
is defined as
\begin{equation}
g(\bmath{x}) = \frac{\gamma(\bmath{x})}{1-\kappa(\bmath{x})},
\label{eq:reduced_shear}
\end{equation}
where the convergence $\kappa(\bmath{x}) =
\Sigma(\bmath{x})/\Sigma_{\rm crit}$ and the shear $\gamma(\bmath{x})$
can, in general, be written as a convolution integral over the
convergence $\kappa(\bmath{x})$ (see
e.g. \citealt{bridle99}). $\Sigma_{\rm crit}$ is the critical surface
mass density
\begin{equation}
\Sigma_{\rm crit} = \frac{c^2}{4 \pi G}\frac{D_{\rm s}}{D_{\rm l} D_{\rm ls}},
\label{eq:sigcrit}
\end{equation}
where $D_{\rm s}$, $D_{\rm l}$ and $D_{\rm ls}$ are the angular-diameter distances between, respectively, the observer and each galaxy, the observer and the
lens, and the lens and each galaxy. In general, the redshifts of each background galaxy can be different, but are assumed to be known. The lensing effect is
weak or strong if $\kappa \ll 1$ or $\kappa \gtrsim 1$ respectively. The observed ellipticity $\epsilon(\bmath{x})$ can be
converted to source ellipticity (i.e. prior to lensing) $\epsilon^{\rm s}(\bmath{x})$ as follows (\citealt{1997A&A...318..687S}):
\begin{equation}
\epsilon^{\rm s}(\bmath{x}) = \left\{ 
\begin{array}{ll}
\frac{\epsilon(\bmath{x})-g(\bmath{x})}{1-g^{*}(\bmath{x})\epsilon(\bmath{x})} & \mbox{$\left( \left| g \right| \le 1 \right)$} \\ & \\
\frac{1-g(\bmath{x})\epsilon^{*}(\bmath{x})}{\epsilon^{*}(\bmath{x})-g^{*}(\bmath{x})} & \mbox{$\left( \left| g \right| > 1 \right)$}
\end{array}
\right.,
\label{eq:ell}
\end{equation}
where an asterisk denotes complex conjugation. The inverse
transformation can be obtained by swapping $\epsilon^{\rm
  s}(\bmath{x})$ with $\epsilon(\bmath{x})$ and by replacing
$g(\bmath{x})$ with $-g(\bmath{x})$.

The number count of the background galaxies is changed by lensing due
to two competing effects: (a) some faint sources in highly magnified
regions are made brighter and pushed over the flux limit, but (b) the
same sources are stretched across the sky reducing the number count of
these sources. The observed number density of sources, $n$ is related
to the actual background number density $n_0$ by the flux limit
$\alpha$ through $n = n_0 \mu^{\alpha - 1}$, where $\mu$ is the
lensing magnification defined by $\mu^{-1} = (1 - \kappa)^2 -
|\gamma|^2$.

\subsection{Triaxial NFW model}\label{sec:lensing:triaxial_NFW}

A model cluster density profile can be determined from numerical $N$-body simulations of large-scale structure formation in a $\Lambda$CDM universe.  In
particular the NFW profile (\citealt{navarro97}) provides a good fit to the simulations. Assuming spherical symmetry, the NFW profile is given by:
\begin{equation}
\rho(r) = \frac{\rho_{\rm s}}{(r/r_{\rm s})(1+r/r_{\rm s})^2},
\label{eq:NFW}
\end{equation}
where $r_{\rm s}$ and $\rho_{\rm s}$ are the radius and density at which the logarithmic slope breaks from $-1$ to $-3$. The extension to a triaxial cluster
is performed simply by defining the triaxial radius $r$, such that
\begin{equation}
r^2 = \frac{X^2}{a^2} + \frac{Y^2}{b^2} + \frac{Z^2}{c^2}, \quad (a \leq b \leq c = 1),
\label{eq:NFW_r}
\end{equation}
where the coordinates $X$, $Y$ and $Z$ lie along the principal axes of the cluster, and $a$, $b$ and $c$ are the semi-minor, semi-intermediate and semi-major
axes respectively of the iso-density ellipsoid $r=1$.  We define $M_{200}$ as the mass contained within the triaxial radius $r=r_{200}$ at which the density
is $200$ times the cosmological critical density $\rho_{\rm crit}$ at the redshift of the cluster. The concentration parameter $C$ is a measure of the halo
concentration and is defined as $C = r_{200}/r_{s}$.

Following \citet{2003ApJ...599....7O}, we define the orientation angles $(\theta,\phi)$ such that they represent the polar angles of the line-of-sight of the
observer, as measured in the principal coordinate system of the triaxial cluster. We set the semi-major axis $c=1$ and therefore the parameters $a$ and $b$
serve as the minor:major and intermediate:major axis ratios respectively. Thus, for a triaxial NFW, cluster parameters are $\mathbf{\Theta} = (x_0, y_0, a,
b, \theta, \phi, M_{200}, C, z)$, where $x_0$ and $y_0$ are the spatial coordinates in the observers $xy$-plane at which the cluster is centred, and $z$ is
its redshift.

The projected surface mass density of a triaxial cluster will have elliptical symmetry and by translating the observer's $xy$-plane such that the cluster
centre is at the origin, and subsequently rotating the coordinate system by the position angle $\psi$ of the iso-surface-density ellipses, the elliptical
radius $\zeta$, in the resulting $(\bar{x},\bar{y})$ coordinate system, can be written as
\begin{equation}
\zeta^{2} = \frac{\bar{x}^{2}}{q_{x}^{2}} + \frac{\bar{y}^{2}}{q_{y}^{2}},
\label{eq:NFW_elliptical_r}
\end{equation}
where $q_x$ and $q_y$ are the semi-major and semi-minor axes, respectively, of the ellipse $\zeta=1$, which corresponds to the projection of the ellipsoid
$r=1$. The position angle $\psi$ is given by
\begin{equation}
\psi = {\textstyle\frac{1}{2}} \arctan\left(\frac{\mathcal{B}}{\mathcal{A}-\mathcal{C}}\right),
\label{eq:NFW_psi}
\end{equation}
and the projected axis ratios read
\begin{eqnarray}
q_{X}^{2} & = & \frac{2f}{\mathcal{A} + \mathcal{C} - \sqrt{\left(\mathcal{A} - \mathcal{C}\right)^{2} +
\mathcal{B}^{2}}},
\label{eq:NFW_qx} \\
q_{Y}^{2} & = & \frac{2f}{\mathcal{A} + \mathcal{C} + \sqrt{\left(\mathcal{A} - \mathcal{C}\right)^{2} +
\mathcal{B}^{2}}},
\label{eq:NFW_qy}
\end{eqnarray}
where
\begin{eqnarray}
f & = & \sin^{2}\theta \left( \frac{c^2}{a^2} \cos^{2}\phi + \frac{c^2}{b^2} \sin^{2}\phi \right) + \cos^{2}\theta,
\label{eq:NFW_f} \\
\mathcal{A} & = & \cos^{2}\theta \left( \frac{c^2}{a^2} \sin^{2}\phi + \frac{c^2}{b^2} \cos^{2}\phi \right) +
\frac{c^2}{a^2} \frac{c^2}{b^2} \sin^{2}\theta,
\label{eq:NFW_A} \\
\mathcal{B} & = & \cos\theta \sin2\phi \left( \frac{c^2}{a^2} - \frac{c^2}{b^2} \right),
\label{eq:NFW_B} \\
\mathcal{C} & = & \frac{c^2}{b^2} \sin^{2}\phi + \frac{c^2}{a^2} \cos^{2}\phi.
\label{eq:NFW_C}
\end{eqnarray}
Note that $q_X \geq q_Y$ for the given rotation angle $\psi$. We define the axis ratio $q$ as $q = q_Y/q_X$, which defines the ellipticities of the projected
iso-density contours of the triaxial cluster.

The lensing shear fields $\bar{\gamma}_1(\bar{x},\bar{y})$ and $\bar{\gamma}_2(\bar{x},\bar{y})$, and convergence $\bar{\kappa}(\bar{x},\bar{y})$ in the
$(\bar{x},\bar{y})$ coordinate system can be calculated as the second derivatives of the lensing potential $\bar{\Phi}(\bar{x},\bar{y}) = \Phi(x,y)$ as
follows (with commas indicating differentiation):
\begin{eqnarray}
\bar{\gamma}_1 & = & \frac{1}{2} \left( \bar{\Phi}_{,\bar{x}\bar{x}} 
- \bar{\Phi}_{,\bar{y}\bar{y}} \right),
\label{eq:NFW_gamma1} \\
\bar{\gamma}_2 & = & \bar{\Phi}_{,\bar{x}\bar{y}},
\label{eq:NFW_gamma2} \\
\bar{\kappa} & = & \frac{1}{2} \left( \bar{\Phi}_{,\bar{x}\bar{x}} 
+ \bar{\Phi}_{,\bar{y}\bar{y}} \right).
\label{eq:NFW_kappa}
\end{eqnarray}
In order to apply the method of \citet{1990AA...231...19S} and \citet{2001astro.ph..2341K} to evaluate these lensing formulae, it is necessary, however, to
perform a further coordinate scaling $\bar{\bar{x}}=\bar{x}/q_x$, $\bar{\bar{y}}=\bar{y}/q_x$ such that the elliptical radius can be written as
\begin{equation}
\zeta^{2} = \bar{\bar{x}}^2+ \frac{\bar{\bar{y}}^{2}}{q^{2}},
\label{eq:NFW_elliptical_keeton_r}
\end{equation}
and so the semi-major axis of the ellipse $\zeta=1$ is equal to unity.

In the $(\bar{\bar{x}},\bar{\bar{y}})$ coordinate system, one can now express the second derivatives of the lensing potential
$\bar{\bar{\Phi}}(\bar{\bar{x}},\bar{\bar{y}})$ as follows\footnote{The expression for the integral $J_{n}(X, Y)$
in Sec. 4 of \citet{2001astro.ph..2341K} should have $\kappa(\zeta(u))$ instead of $\kappa(\zeta(u)^2)$, in the
numerator.}:
\begin{eqnarray}
\bar{\bar{\Phi}}_{,\bar{\bar{x}}\bar{\bar{x}}}(\bar{\bar{x}},\bar{\bar{y}})
& = & q \bar{\bar{x}}^2 K_{0}(\bar{\bar{x}},\bar{\bar{y}}) + q
J_{0}(\bar{\bar{x}},\bar{\bar{y}}),
\label{eq:NFW_phiXX} \\
\bar{\bar{\Phi}}_{,\bar{\bar{y}}\bar{\bar{y}}}(\bar{\bar{x}},\bar{\bar{y}})
& = & q \bar{\bar{y}}^2 K_{2}(\bar{\bar{x}},\bar{\bar{y}}) + q
J_{1}(\bar{\bar{x}},\bar{\bar{y}}),
\label{eq:NFW_phiYY} \\
\bar{\bar{\Phi}}_{,\bar{\bar{x}}\bar{\bar{y}}}(\bar{\bar{x}},\bar{\bar{y}})
& = & q \bar{\bar{x}}\bar{\bar{y}} K_{1}(\bar{\bar{x}},\bar{\bar{y}}),
\label{eq:NFW_phiXY}
\end{eqnarray}
where
\begin{eqnarray}
K_{n}(\bar{\bar{x}},\bar{\bar{y}}) & = & \int_{0}^{1} \frac{u \kappa^{\prime}\left( \zeta\left( u \right) \right)}
{\zeta\left( u \right) \left[ 1 - \left( 1 - q^{2} \right) u \right]^{n + 1/2}} \mathrm{d}u,
\label{eq:NFW_K} \\
J_{n}(\bar{\bar{x}},\bar{\bar{y}}) & = & \int_{0}^{1} \frac{\kappa\left( \zeta\left( u \right) \right)}{\left[ 1
- \left( 1 - q^{2} \right) u \right]^{n + 1/2}} \mathrm{d}u,
\label{eq:NFW_J}
\end{eqnarray}
in which
\begin{equation}
\zeta\left(u\right)^{2} = u \left( \bar{\bar{x}}^{2} + \frac{\bar{\bar{y}}^{2}}{1 - \left( 1 - q^{2} \right) u} \right),
\label{eq:NFW_zeta}
\end{equation}
and
\begin{equation}
\kappa^{\prime}\left( \zeta\left(u\right) \right) = \frac{d\kappa\left( \zeta\left(u\right)
\right)}{d\zeta\left(u\right)}.
\label{eq:NFW_kappaprime}
\end{equation}
The surface mass density $\kappa\left(\zeta\right)$ in the above equations can be expressed in terms of the spherical surface mass density
$\kappa_s\left(\zeta\right)$ (see \citealt{bartelmann96} for the expression for $\kappa_s\left(\zeta\right)$ for the NFW profile) as follows (see
\citealt{2003ApJ...599....7O} for the derivation):
\begin{equation}
\kappa\left(\zeta\right) = \frac{\kappa_s\left(\zeta\right)}{\sqrt{f}},
\label{eq:triNFW_kappa}
\end{equation}
where $f$ is given in Eq.~(\ref{eq:NFW_f}).

The shear and convergence fields in the $(\bar{x},\bar{y})$ coordinates, given in Eqs.~(\ref{eq:NFW_gamma1})--(\ref{eq:NFW_kappa}) are straigtforwardly
calculated from the derivatives of the lensing potentials Eqs.~(\ref{eq:NFW_phiXX})--(\ref{eq:NFW_phiXY}) using the fact that
$\bar{\Phi}_{,\bar{x}\bar{x}}(\bar{x},\bar{y}) = \bar{\bar{\Phi}}_{,\bar{\bar{x}}\bar{\bar{x}}}(\bar{\bar{x}},\bar{\bar{y}})$, and similarly for the other
terms (since the gravitational potential has a scaling dimension of 2). Finally, the convergence and shear values in the observer's original $(x,y)$
coordinate system (shifted such that the cluster centre is at the origin) are obtained by rotating the shear vector by angle $\psi$, given in
Eq.~(\ref{eq:NFW_psi}), as follows\footnote{Although the description of triaxial lensing in \citet{corless07} and \citet{corless08} makes no mention of the successive
coordinate rescaling and rotation discussed above, \citet{CorlessThesis} does mention that all triaxial projections are rotated by
their position angle $\psi$ before lensing integrals are calculated and that the resulting shear and convergence values are rotated back to the
original orientation of the lens. Also, Eq.~(A18) in \citet{corless08} should not have $q_X$ in the denominator. However, the correct form of equations
were used in their analysis code (King, private communication).}
\begin{eqnarray}
\kappa(x,y) & = & \bar{\kappa}(\bar{x},\bar{y}),\label{eq:NFW_K_XY} \\
\gamma_1(x,y) & = & |\bar{\gamma}(\bar{x},\bar{y})|\cos\left(2\left(\alpha+\psi\right)\right),\label{eq:NFW_gamma1_XY} \\
\gamma_2(x,y) & = & |\bar{\gamma}(\bar{x},\bar{y})|\sin\left(2\left(\alpha+\psi\right)\right),\label{eq:NFW_gamma2_XY}
\end{eqnarray}
where
\begin{equation}
|\bar{\gamma}(\bar{x},\bar{y})| = \sqrt{\bar{\gamma}_{1}^{2}(\bar{x},\bar{y}) +
\bar{\gamma}_{2}^{2}(\bar{x},\bar{y})},
\label{eq:triNFW_modgamma}
\end{equation}
and
\begin{equation}
\alpha = \frac{1}{2} \arctan\left( \frac{\bar{\gamma}_2(\bar{x},\bar{y})}
{\bar{\gamma}_1(\bar{x},\bar{y})}\right).
\label{eq:triNFW_phigamma}
\end{equation}

We have tested the triaxial NFW lensing equations given above by numerically integrating the mass density of a given triaxial cluster, along the
line-of-sight to calculate the surface mass density $\kappa(x,y)$ which was found to be equal to the $\kappa(x,y)$ calculated by using
Eqs.~(\ref{eq:NFW_kappa}) and ~(\ref{eq:triNFW_kappa}).

\subsection{Weak lensing likelihood}\label{sec:lensing:likelihood}

%
\begin{table*}
\begin{center}
\begin{tabular}{cccc}
\hline
Parameters & Flat Priors & Bett Priors & Spherical Priors \\
\hline
$x_0$ & $-12^{\prime} \leq x_0 \leq 12^{\prime}$ & $-12^{\prime} \leq x_0 \leq 12^{\prime}$ & $-12^{\prime} \leq x_0 \leq 12^{\prime}$ \\
$y_0$ & $-12^{\prime} \leq y_0 \leq 12^{\prime}$ & $-12^{\prime} \leq y_0 \leq 12^{\prime}$ & $-12^{\prime} \leq y_0 \leq 12^{\prime}$ \\
$a/b$ & $0.2 \leq a/b \leq 1.0$ & $0.800 \pm 0.125$ & $a/b = 1$ \\
$b$ & $0.2 \leq b \leq 1.0$ & $0.85 \pm 0.10$ & $b = 1$ \\
$\theta/\mbox{rad}$ & $0 \leq \sin(\theta) \leq 1$ & $0 \leq \sin(\theta) \leq 1$ & $\theta = 0$ \\
$\phi/\mbox{rad}$ & $0 \leq \phi \leq 2\pi$ & $0 \leq \phi \leq 2\pi$ & $\phi = 0$ \\
$M_{200}/h^{-1} M_{\sun}$ & $14.0 \leq \log_{10}(M_{200}) \leq 16$ & $14.0 \leq \log_{10}(M_{200}) \leq 16$ & $14.0 \leq \log_{10}(M_{200}) \leq 16$ \\
$C$ & $0 \leq C \leq 15$ & $0 \leq C \leq 15$ & $0 \leq C \leq 15$ \\
$z$ & $z = 0.2$ & $z = 0.2$ & $z = 0.2$ \\
\hline
\end{tabular}
\caption{Priors for the cluster parameters. Inequalities denote
  uniform prior probability between the given limits, equalities mean
  delta function priors, and $(a \pm b)$ denotes a Gaussian prior with
  mean $a$ and variance $b^2$.}
\label{tab:cluster_priors}
\end{center}
\end{table*}

The observed complex ellipticity components of the $N_{\rm gal}$ background galaxies can be ordered into a data
vector $\bmath{d}$ with components
\begin{equation}
d_i = \left\{ 
\begin{array}{ll}
\textrm{Re}(\epsilon_{i}) & \mbox{$\left(i \leq N_{\rm gal}\right)$} \\ & \\
\textrm{Im}(\epsilon_{i-N_{\rm gal}}) & \mbox{$\left(N_{\rm gal}+1 \leq i \leq 2N_{\rm gal}\right)$}
\end{array}
\right.
.
\label{eq:gldvect}
\end{equation}
Likewise the corresponding components of the complex reduced shear
$g(\bmath{x}_i)$ at each galaxy position, as predicted by the cluster
model, can be arranged into the predicted data vector $\bmath{d}^{\rm
  P}$, with the arrangement of components matching (\ref{eq:gldvect}).

The uncertainty on the components (i.e. real and imaginary parts) of
the measured ellipticity $\epsilon$ consists of two contributions. The
components of the intrinsic ellipticity, $\epsilon_{\rm s}$, of the
background galaxies (i.e. prior to lensing) may be taken as having
been drawn independently from a Gaussian distribution with mean zero
and variance $\sigma_{\rm int}^2$. If we denote distribution of 
$\epsilon_{\rm s}$ by $p_{\rm int}(\epsilon_{\rm s})$, the distribution of the
lensed ellipticities $p_{\epsilon_\ell}(\epsilon_\ell)$ is then
given by:
\begin{equation}
p_{\epsilon_\ell}(\epsilon_\ell) = p_{\rm int}(\epsilon_{\rm s}) \left| \frac{\mathrm{d}\epsilon_{\rm s}}{\mathrm{d}\epsilon_\ell}
\right|^2,
\label{eq:ell_dist}
\end{equation}
where we have taken account of $\epsilon_{\rm s}$ and $\epsilon_\ell$
both being complex variables. The Jacobian determinant can be
found using Eq.\eqref{eq:ell} and is given by:
\begin{equation}
\left| \frac{\mathrm{d}\epsilon_{\rm s}}{\mathrm{d}\epsilon_\ell} \right| = \left\{ 
\begin{array}{ll}
\frac{1 - \left| g \right|^2}{\left| \epsilon_\ell g^{*} - 1 \right|^2} & \mbox{$\left( \left| g \right| \le 1 \right)$} \\ & \\
\frac{\left| g \right|^2 - 1}{\left| \epsilon_\ell - g \right|^2} & \mbox{$\left( \left| g \right| > 1 \right)$}
\end{array}
\right.
.
\label{eq:jacobian}
\end{equation}

The second source of uncertainty on the measured ellipticity is due to
the errors introduced by the galaxy shape estimation procedure, such
that $\epsilon=\epsilon_\ell+\Delta\epsilon_\ell$, where the
components of $\Delta\epsilon_\ell$ can be modelled as drawn from a
Gaussian with mean zero and variance $\sigma^2_{\rm obs}$. Denoting
this distribution by $p_{\rm obs}(\Delta\epsilon_\ell)$, the
probability distribution for the measured ellipticities can be
obtained as follows:
\begin{equation}
p_{\epsilon}(\epsilon) = \int{p_{\epsilon_\ell}(\epsilon_\ell)p_{{\rm obs}}(\epsilon - \epsilon_\ell)}\mathrm{d}\epsilon_\ell.
\label{eq:convolution}
\end{equation}

If the reduced shear is small i.e. for $|g| \ll 1$, the Jacobian determinant in Eq.~\eqref{eq:jacobian} reduces to
unity and we have $p_{\epsilon_\ell}(\epsilon_\ell) 
= p_{\rm int}(\epsilon_{\rm s})$. This reduces
Eq.~\eqref{eq:convolution} to a convolution of two Gaussians with means zero and standard deviations $\sigma_{\rm
int}$ and $\sigma_{\rm obs}$ respectively. Consequently $p_{\epsilon}(\epsilon)$ is a Gaussian with mean zero and
standard deviation $\sigma_{\rm i}$ corresponding to the $i$th galaxy, given by:
\begin{equation}
\sigma^2_{i} = \sigma^2_{\rm obs}+\sigma^2_{\rm int}.
\label{eq:newsigma}
\end{equation}
This leads to a diagonal noise covariance matrix $\mathbf{C}$ on the ellipticity components. This is however, not
the case when $|g| \simeq 1$. Since the expected reduced shear from the low redshift simulated clusters in this study is not too large, we adopt this approximation.

%
%

As shown by \citet{marshall03}, we can then write the likelihood function as
$\mathcal{L}(\mathbf{\Theta}) = Z_L^{-1} \exp (-{\textstyle\frac{1}{2}}\chi^2)$,
where $\chi^2$ is the usual misfit statistic 
\begin{equation}
\chi^2 
= ( \bmath{d} - \bmath{d}^{\rm P})^{\rm T}
\mathbf{C}^{-1} 
( \bmath{d} - 
\bmath{d}^{\rm P})\\
= \sum_{i=1}^{N_{\rm gal}} \sum_{j=1}^{2} \frac{\left(\epsilon_{j,i} -
g_{j}(\bmath{x}_i)\right)^2}{\sigma_{\rm i}^2},
\label{eq:glchisq}
\end{equation}
and the normalisation factor is $Z_{L} = (2 \pi)^{2N_{\rm gal}/2}
|\mathbf{C}|^{1/2}$.  Note that is it necessary to include this
normalisation factor in the likelihood, since the covariance matrix
$\mathbf{C}$ is not constant, but depends on the cluster model
parameters through the predicted shear terms in (\ref{eq:newsigma}).

\subsection{Priors on cluster parameters}\label{sec:application:priors}

To determine the model completely it only remains to specify the prior $\pi(\mathbf{\Theta})$ on the cluster parameters $\mathbf{\Theta} =
(x_0, y_0, a, b, \theta, \phi, M_{\rm 200}, C, z)$. The choice of prior is particularly important for weak lensing analysis since the problem
is inherently underconstrained and therefore any prior information available about the cluster parameter is extremely useful. One should
however be careful in the choice of priors not to impose too strong assumptions which may lead to erroneous inferences.

We use three different priors which we call flat, Bett and spherical. These priors are specified in Tab.~\ref{tab:cluster_priors}. In all 3
cases, we assume that cluster redshift is known. It should also be noted that although the orientation angles are defined over $0 < \theta <
\pi$ and $0 < \phi < 2 \pi$, because of the elliptical nature of the projected density contours, unique lensing profiles are possible only over
the range $0 < \theta < \pi / 2$ and $0 < \phi < \pi$. Apart from the spherical case, the prior on $\theta$ is uniform in $\sin(\theta)$.

The most restrictive case is that of the spherical priors, which assume the axis ratios to be fixed $a = b = 1$,
and hence correspond to a spherical cluster model.  This is clearly a very strong assumption and there is a
debate in the literature whether this assumption is behind the deviations from $\Lambda$CDM of the observed
gravitational lensing results (see e.g. \citet{corless07, 2005ApJ...632..841O, 2005A&A...443..793G}). The least
restrictive case is given by the flat priors, for which assume a uniform prior distribution on the axis ratios
with $0.2 \leq b \leq 1.0$ and $0.2 \leq a/b \leq 1.0$. An intermediate choice is to use the distribution of axis
ratios derived from $N$-body simulations as the prior on axis ratios. We therefore also consider the Bett priors
specified in \citet{corless08}, for which the distribution of axis ratios is taken from $N$-body simulations by
\citet{bett07}. Bett priors assume a Gaussian distribution on the axis ratios $a/b$ and $b$ with means and
standard deviations given in Tab.~\ref{tab:cluster_priors}.

\begin{figure*}
\begin{center}
\subfigure[]{\includegraphics[width=0.8\columnwidth]{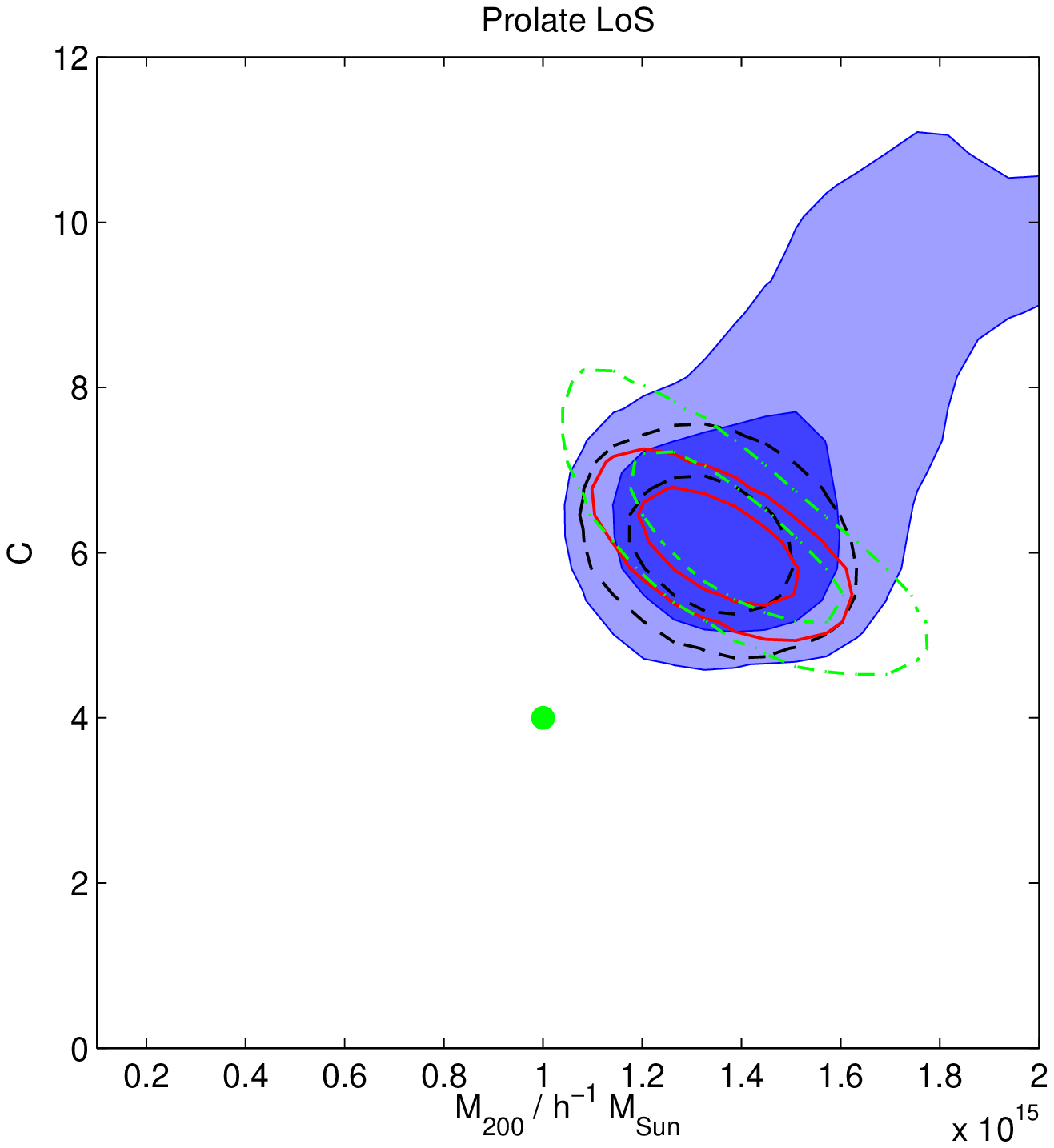}}\hspace{0.3cm}
\subfigure[]{\includegraphics[width=0.8\columnwidth]{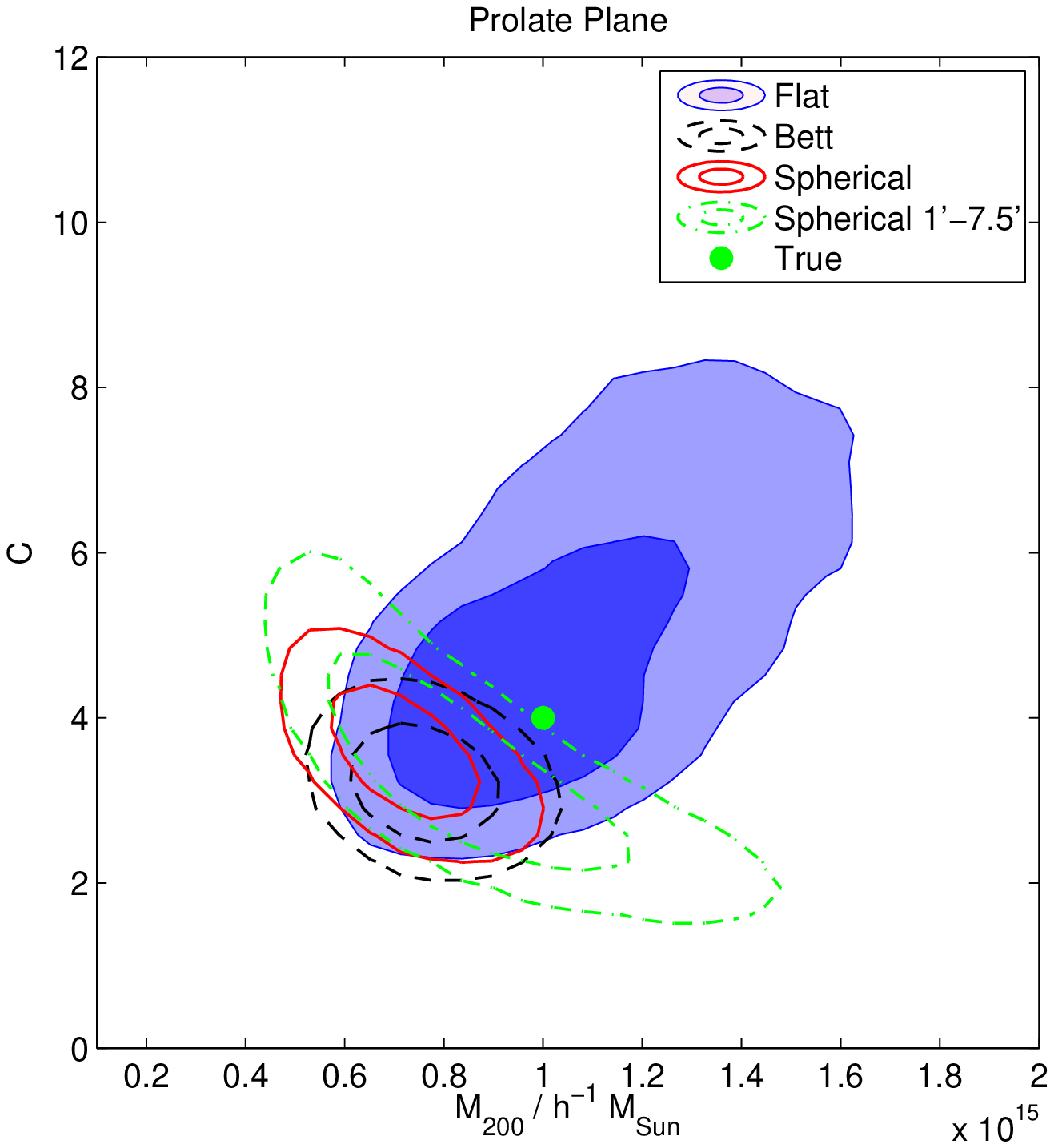}}\hspace{0.3cm}
\subfigure[]{\includegraphics[width=0.8\columnwidth]{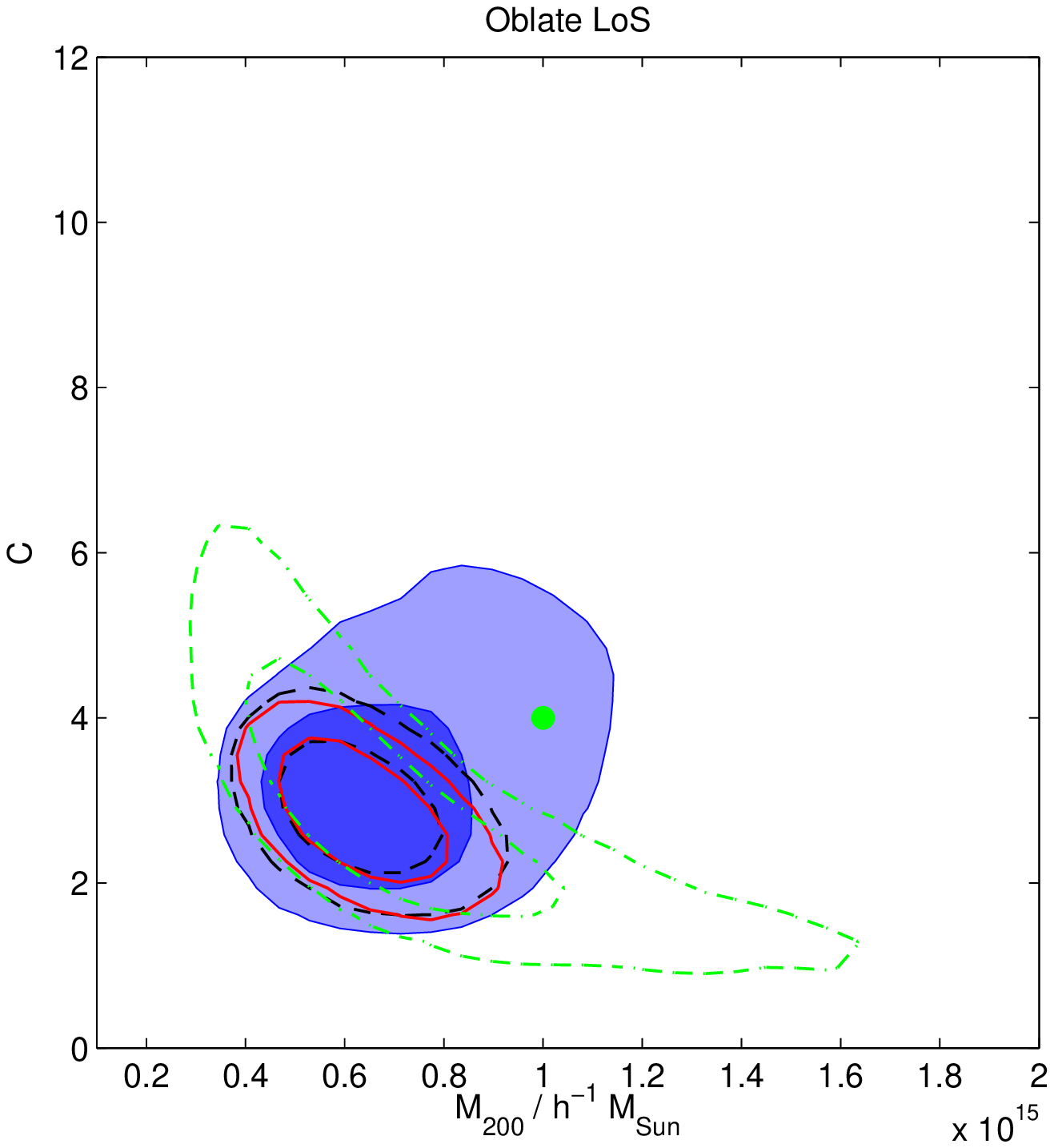}}\hspace{0.3cm}
\subfigure[]{\includegraphics[width=0.8\columnwidth]{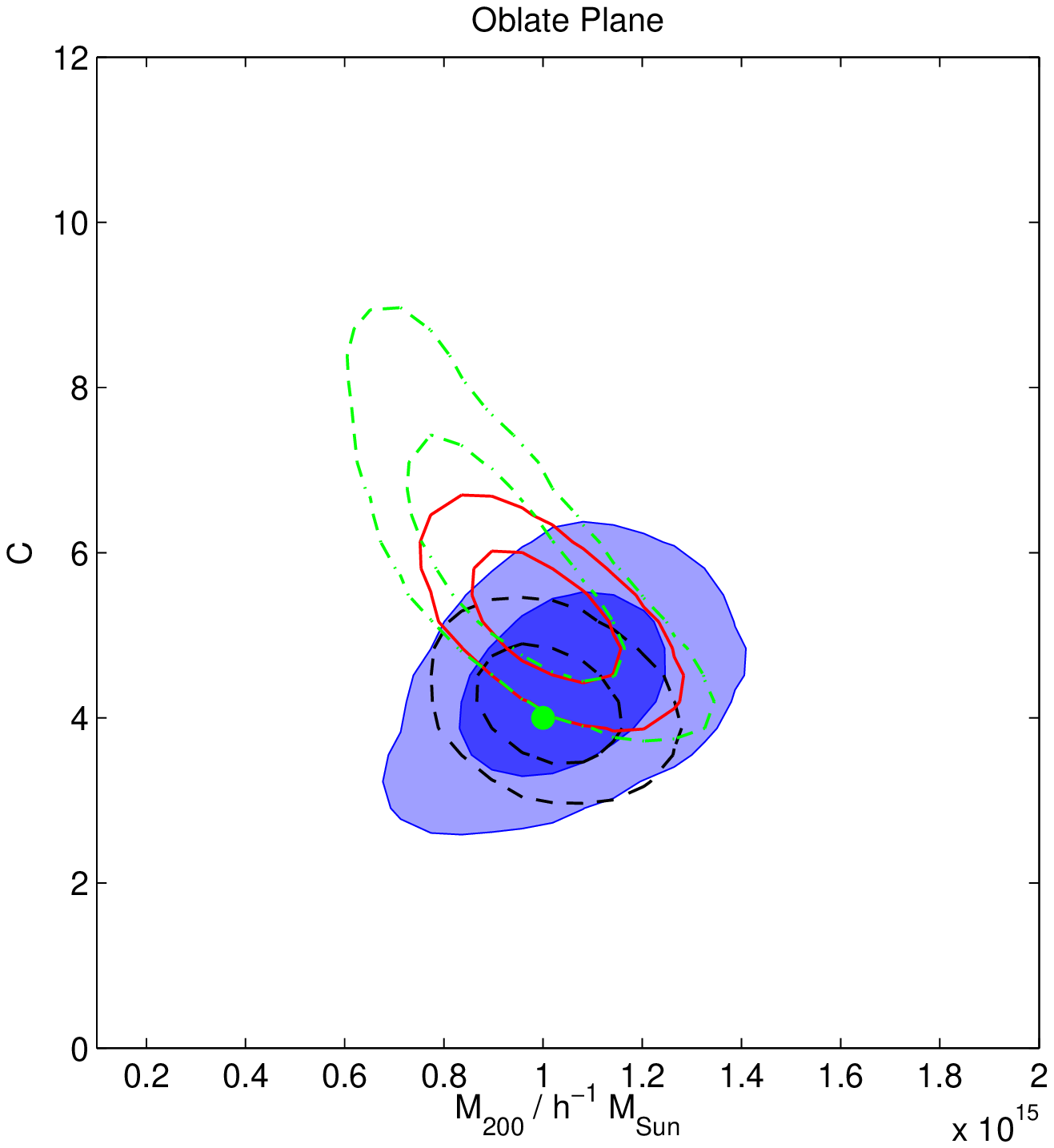}}\hspace{0.3cm}
\caption{2-D marginalized posterior probability distributions in the
  $M_{\rm 200}$-$C$ plane for (a) prolate LoS cluster, (b) prolate
  plane cluster, (c) oblate LoS cluster and (d) oblate plane
  cluster. The contours delineate the 68 and 95 per cent confidence
  regions, respectively, for the different assumed
  priors.\label{fig:triaxial_2D}}
\end{center}
\end{figure*}
%

\section{Application to simulated observations}\label{sec:application}

\subsection{Simulations}\label{sec:application:simulations}

Following \citet{corless08}, we simulate prolate ($a = b = 0.4$, i.e. `cigar-shaped') and oblate ($a = 0.4$, $b = 1$, i.e. `pancake-shaped') clusters in two
orientations: (a) Line-of-sight (LoS): with the `odd' axis along the line of sight ($\theta = 0$ for prolate and $\theta = \pi/2, \phi = 0$ for oblate), and
(b) Plane: with the `odd' axis in the plane of the sky ($\theta = \pi/2$ for prolate and $\theta = \phi = 0$ for oblate). LoS orientation results in
spherical projected iso-density contours for our chosen prolate and oblate clusters. In Plane orientation, the projected iso-density contours are elliptical
with a position angle $\psi$ of $\pi/2$ and $0$, respectively, for prolate and oblate clusters. We also simulate a spherically symmetric cluster ($a = b =
1$) to determine the effects of fitting an (overly complicated) triaxial model to it.

All the simulated clusters were assumed to have the NFW density profile as given in Eq.~(\ref{eq:NFW}), with $x_0 = 0$, $y_0 = 0$, $M_{\rm 200} = 10^{15}
h^{-1} M_{\sun}$, $C = 4$ and $z = 0.2$. The simulations were carried out for a field $12^{\prime}$ in radius, with background source density $n_0 =
30/$arcmin$^2$, typical of ground-based observations. We assume all the background galaxies lie at redshift $z = 1$, this is justified by the low redshift
($z = 0.2$) of the simulated lens. The galaxies are positioned randomly in the field with their shapes drawn from a Gaussian distribution with standard
deviation $\sigma_{\epsilon} = \sqrt{\sigma_{\rm int}^2 + \sigma_{\rm obs}^2}$, where $\sigma_{\rm int}$ and $\sigma_{\rm obs}$ are the intrinsic and
observational dispersion of galaxy shapes as discussed in Sec.~\ref{sec:lensing:likelihood}. We set $\sigma_{\rm int} = 0.2$ and $\sigma_{\rm obs} = 0.1$.
These galaxies are then lensed by the cluster as described in Sec.~\ref{sec:lensing:background}. The number density of background galaxies is then reduced
according to $n = n_0 \mu^{\alpha - 1}$ with the flux limit $\alpha = 0.5$, as in \citet{1997A&A...321..353F} and \citet{corless08}. Throughout we assume a
concordance $\Lambda$CDM cosmology ($\Omega_{\rm m,0}=0.3$, $\Omega_{\Lambda,0}=0.7$ and $h=0.7$).

\subsection{Analysis and Results}\label{sec:application:results}

\begin{figure}
\begin{center}
\includegraphics[width=0.8\columnwidth]{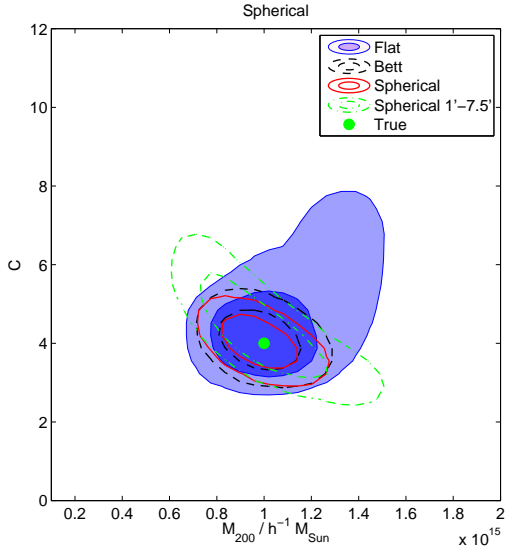}
\caption{2-D marginalized posterior probability distributions in the
$M_{\rm 200}$-$C$ plane for a spherical cluster.}
\label{fig:spherical_2D}
\end{center}
\end{figure}

Fig.~\ref{fig:triaxial_2D} shows the posterior probability distributions, marginalized in the $M_{\rm 200}$-$C$
plane, obtained using the different priors discussed in Sec.~\ref{sec:application:priors}, for fitting a triaxial
NFW model to the triaxial cluster simulations discussed in Sec.~\ref{sec:application:simulations}.  We also
analyse our simulated data-sets assuming the spherical model and include the results in
Fig.~\ref{fig:triaxial_2D}. To emulate the analysis of \citet{corless07}, for the spherical model we further
consider the case where one removes the galaxies within $1^{\prime}$ of the cluster centre to avoid the strong
lensing regime and only simulate galaxies out to $7.5^{\prime}$ from the cluster centre. The results for fitting
a triaxial NFW model to a spherical cluster are shown in Fig.~\ref{fig:spherical_2D}.

As expected, in all cases the flat priors result in broadest posterior distributions, while the spherical priors give the most constrained posteriors.  For
the flat priors, we see from Fig.~\ref{fig:triaxial_2D} that each posterior distribution contains the true parameter values with the 95 per cent confidence
region, except for the prolate LoS cluster which has significantly overestimated values for both $M_{\rm 200}$ and $C$.  In LoS orientation, the surface mass
density of the cluster has spherical symmetry and, coupled with the thin lens approximation, one would expect the axis ratios and orientation angles to be
largely unconstrained by the observed galaxy ellipticities. Therefore, the inferences on axis ratios $a, b$ and orientation angles $\theta, \phi$ for our LoS
prolate and oblate clusters are expected to be driven by their prior distributions. The overestimation of $M_{\rm 200}$ and $C$ for the prolate LoS cluster
occurs because it has $\theta = 0$ and $a/b = 1$, which is highly disfavoured by the prior distributions employed.  In the absence of strong constraints from
the data, the priors thus pull these parameters away from their true values. The consequent inaccuracies in the estimates of $\theta$ and $a/b$ result in an
overestimation of $M_{\rm 200}$ and $C$. We found that fixing the axis ratios $a$ and $b$ or the orientation angle $\theta$ to their true values removes the
bias in the estimation of $M_{\rm 200}$ and $C$. The oblate LoS cluster has $\theta = \pi/2, \phi = 0$ which is not disfavoured by the prior distribution and
therefore we do not see a significant bias in the posterior distributions. For the prolate plane and oblate LoS orientations, it is worth noting that the
degeneracy directions of the posteriors with flat priors are almost orthogonal to those obtained in these cases by \citet{corless08} which may be due to the
wider priors we used on the axis ratios, whereas the shapes of the posteriors agree well in the other two cases.

As the priors become more restrictive, the true parameter values do begin to fall noticeably outside the two-dimensional 95 per cent confidence region for
all cases, except the oblate plane cluster. Even for the spherical priors, however, one obtains largely unbiased estimates for $C$, except in the prolate LoS
orientation for which $C$ is significantly overestimated (as explained above). Turning to $M_{\rm 200}$ derived from spherical priors, we see that it is
significantly overestimated in the prolate LoS orientation, noticeably underestimated in the oblate LoS case, slightly underestimated in the prolate plane
orientation, and largely unbiased for the oblate plane cluster.

The most notable inaccuracies in fitting a spherical NFW model are observed when we allow for more than one spherical cluster in the field. Bayesian model
selection can then be used to determine whether multiple spherical clusters provide a better fit to the data. Table ~\ref{tab:cluster_evidence} lists the
Bayesian evidences calculated by our cluster finding algorithm for one triaxial NFW cluster, one spherical NFW cluster and two spherical NFW clusters. As
expected, one triaxial NFW cluster model is preferred for the triaxial cluster simulations with their odd axis in the plane of the sky, which results in the
surface mass density having elliptical contours. LoS orientation, however, results in spherical projected iso-density contours, as discussed in
Sec.~\ref{sec:application:simulations}, and since the prior on $\theta$ completely rules out the true value for the prolate LoS cluster, there is weak
preference for the spherical model over a triaxial cluster model. There is no such mismatch between the true parameter values and the corresponding prior
distributions for the oblate LoS cluster and therefore the Bayesian evidence does not prefer either the spherical or the triaxial model over the other.

\begin{figure*}
\begin{center}
\subfigure[]{\includegraphics[width=0.8\columnwidth]{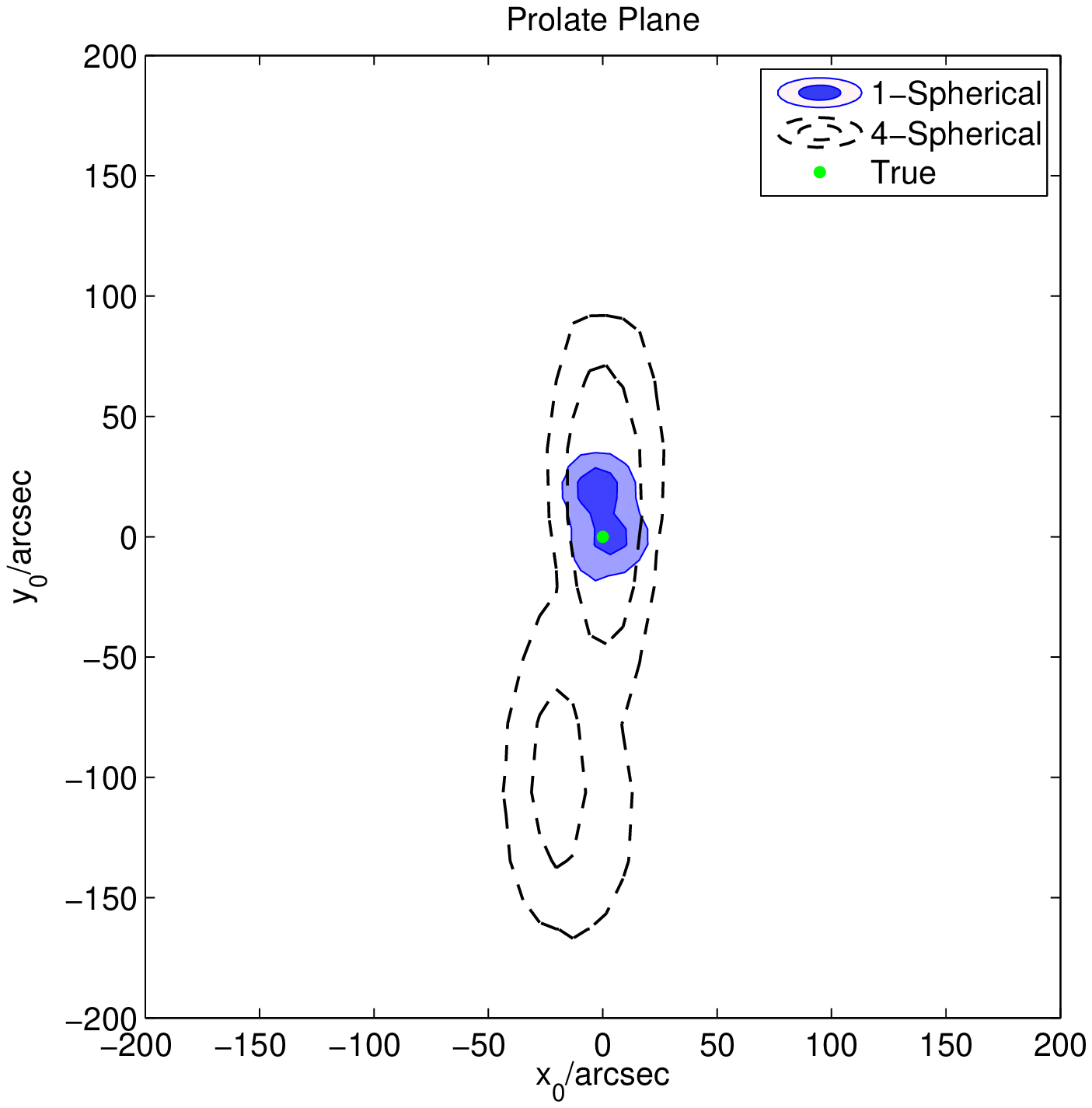}}\hspace{0.3cm}
\subfigure[]{\includegraphics[width=0.8\columnwidth]{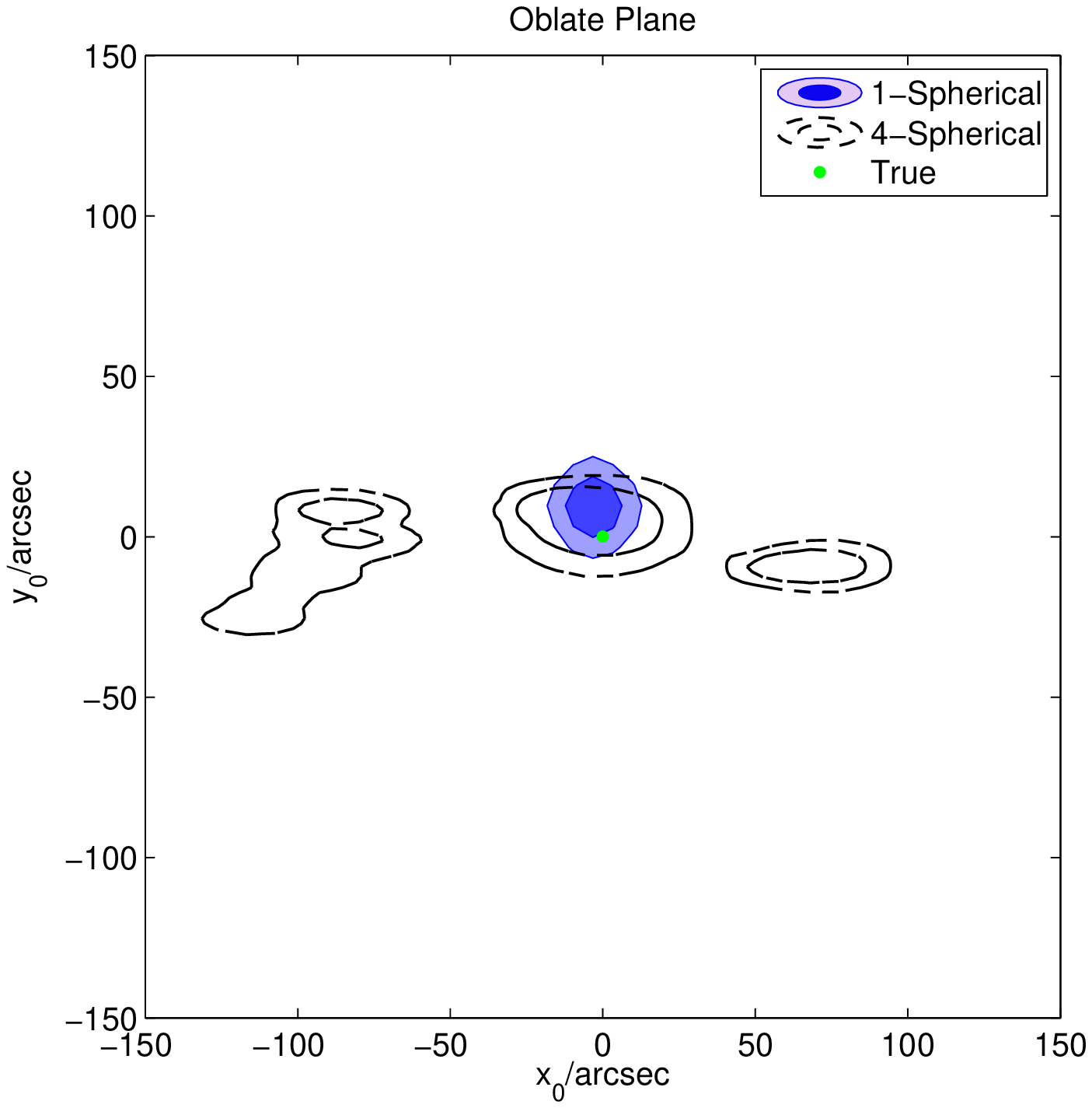}}\hspace{0.3cm}
\caption{2-D marginalized posterior probability distributions in the $x_0$-$y_0$ plane for (a) prolate plane
  cluster, (b) oblate plane cluster, when analysed assuming 1 and 4 spherical cluster models.}
\label{fig:triaxial_multisph_2D}
\end{center}
\end{figure*}

In the case of the triaxial cluster simulations with odd axis in the plane of the sky, and hence elliptical iso-density contours, one would expect multiple
spherical clusters to provide a better fit than just a single spherical cluster. This is confirmed by the evidence values as listed in Table
~\ref{tab:cluster_evidence}. In fact, we fitted up to 4 spherical clusters for both prolate and oblate plane clusters and found the evidence value to be
still increasing. We show the posterior distributions for the spatial coordinates at which the cluster is centred for the 1-spherical and 4-spherical model
fits to the prolate and oblate plane clusters in Fig.~\ref{fig:triaxial_multisph_2D}. One can see multiple clusters aligned along the $x$-axis and $y$-axis,
respectively, for the oblate and prolate plane clusters, which corresponds to the position angle of the projected iso-density contours in each case. It
is clear from this figure that by fitting spherical NFW models to highly triaxial simulations, one would incorrectly infer the presence of multiple clusters.
This is perhaps the biggest drawback of fitting spherical NFW models to intrinsically triaxial clusters and the most important result of this study. It
should also be noted that the single spherical cluster model is favoured in the analysis of the spherical NFW cluster simulation.

\begin{table}
\begin{center}
\begin{tabular}{lrrr}
\hline
Cluster & Triaxial & 1-Spherical & 2-Spherical \\
\hline
Prolate LoS 	& $636.1 \pm 0.3$ & $638.2 \pm 0.3$ & $632.6 \pm 0.3$ \\
Prolate Plane 	& $204.5 \pm 0.3$ & $163.4 \pm 0.2$ & $170.4 \pm 0.3$ \\
Oblate LoS 	& $111.8 \pm 0.3$ & $112.5 \pm 0.3$ & $107.6 \pm 0.3$ \\
Oblate Plane 	& $406.1 \pm 0.3$ & $346.7 \pm 0.2$ & $349.8 \pm 0.3$ \\
Spherical 	& $280.3 \pm 0.2$ & $280.6 \pm 0.2$ & $273.6 \pm 0.3$ \\
\hline
\end{tabular}
\caption{$\log\mathcal{Z}$ values for fitting prolate LoS, prolate plane, oblate LoS, oblate plane and spherical
clusters with triaxial (flat priors), one spherical cluster and two spherical cluster models.}
\label{tab:cluster_evidence}
\end{center}
\end{table}
%

\section{Conclusions}\label{sec:conclusions}

We have presented the weak lensing equations for the triaxial NFW profile. We also presented a method to estimate
the cluster parameters assuming a triaxial NFW model. By considering four highly triaxial NFW galaxy clusters, we
found that, spherical assumptions resulted in significant biases in parameter estimates for the cluster
concentration $C$ only for the prolate cluster in line-of-sight orientation. The cluster mass $M_{\rm 200}$ was
significantly biased, however, being overestimated for triaxial clusters oriented such that there is a lot of
mass along the line-of-sight and underestimated for triaxial clusters with very little mass along the
line-of-sight. The tension between the concentration $C \sim 4$ of clusters predicted in the $\Lambda$CDM model
and the high values $C \sim 8$ estimated in some clusters when assuming a spherical model can therefore be
explained if these clusters have very extreme axis ratios and oriented such that there is a lot of mass along the
line-of-sight. Even more worryingly, the spherical model can lead to multiple cluster detections when applied to
highly triaxial clusters with their odd axis in the plane of the sky. It is therefore important to be cautious in
drawing conclusions about presence of substructure in weak gravitational lensing cluster studies in which a
spherical cluster model has been used.

Even though our analysis algorithm is very efficient, fitting for triaxial cluster still takes a few days
because, unlike the spherical case, the shear equations for triaxial NFW described in
Sec.~\ref{sec:lensing:triaxial_NFW} do not have an analytical solution and need to be solved numerically. This
makes the application of triaxial NFW model for searching large survey fields along the lines of the method
presented in \citet{2008arXiv0810.0781F} unfeasible. In a future study, we plan to train an artificial neural
network for solving these triaxial NFW equations and apply the resultant method on survey fields.

Another important area for studying cluster physics is through coherent analysis of weak/strong lensing, optical,
X-ray and Sunyaev--Zel'dovich (SZ) observations. As we can see, analysis using weak lensing data only results in
large uncertainties in cluster parameters due to inherent underconstrained nature of the problem. It is therefore
important to perform a joint analysis using several different data-sets in order to obtain better constraints on
cluster parameters. \cite{marshall03} presented a method for the joint analysis of lensing and SZ observations
which was further extended in \citet{2009MNRAS.398.2049F} and \citet{2011arXiv1101.5912H}. These studies assumed
a spherical NFW model. We plan to extend this to a triaxial NFW model and incorporate the X-ray observations as
well in a future work.

\section*{Acknowledgements}

We thank Lindsay King and James Mead for useful discussions regarding the triaxial NFW model lensing equations. This work was carried out largely on the {\sc
Cosmos} UK National Cosmology Supercomputer at DAMTP, Cambridge and the Darwin Supercomputer of the University of Cambridge High Performance Computing
Service ({\tt http://www.hpc.cam.ac.uk/}), provided by Dell Inc. using Strategic Research Infrastructure Funding from the Higher Education Funding Council
for England. FF is supported by a Research Fellowship from Trinity Hall, Cambridge.

\bibliographystyle{mn2e}
\bibliography{references}

\label{lastpage}

\end{document}